\documentclass[twocolumn]{aastex631}

\usepackage{newtxtext,newtxmath}

\usepackage[T1]{fontenc}
\usepackage{ae,aecompl}

\usepackage{graphicx}    
\usepackage{amsmath}    
\usepackage{enumerate}
\usepackage{hyperref}

\newcommand{\aref}[1]{\hyperref[#1]{Appendix~\ref{#1}}}

\hypersetup{linkcolor=red}          
\shorttitle{H\textsc{i} deficit of slow-rotating galaxies}
\shortauthors{Wang \& Peng}
\graphicspath{{./}{figures/}}

\begin{document}

\title{The kinematic bimodality: Efficient feedback and cold gas deficiency in slow-rotating galaxies}

\author[0000-0002-6137-6007]{Bitao Wang}
\affiliation{Kavli Institute for Astronomy and Astrophysics, Peking University, Beijing 100871, China}

\author{Yingjie Peng}
\affiliation{Kavli Institute for Astronomy and Astrophysics, Peking University, Beijing 100871, China}
\affiliation{Department of Astronomy, School of Physics, Peking University, Beijing 100871, China}



\begin{abstract}

  The bimodality in the stellar spin of low redshift (massive) galaxies, ubiquitously existing at all star formation levels and in diverse environment, suggests that galaxies grow and quench through two diverged evolutionary pathways.
  For spheroid-dominated galaxies of slow stellar rotation, the age composition and metallicity of their stellar populations evidence a fast quenching history with significant gas outflows.
  In this work, we measure the spin parameter $\lambda_{R_{\rm e}}$, i.e. the normalized specific angular momentum of stars, out of the MaNGA integral field spectroscopy for about 10000 galaxies.
  Among the two thirds with H\textsc{i} follow-up observations ($z\lesssim0.05$), we find that, compared to fast-rotating galaxies of the same stellar mass and star formation, the galaxy population of slower rotation are generally more H\textsc{i} gas-poor, robust against further environmental restriction and with non-detections taken into proper account using stacking technique.
  This cold gas deficit of slow-rotating galaxies is most apparent at high mass $\sim10^{11}\mathcal{M}_{\odot}$ below the star formation main sequence, supporting the pivotal role of gas outflows in their quenching history.
  With hints from H\textsc{i} velocity distributions, we suspect that massive gas outflows among the slow-rotating population are facilitated by high ejective feedback efficiency, which is a result of extensive coupling between disturbed volume-filling cold gas and (commonly) biconical feedback from central black holes.
  By contrast, in fast-rotating disc galaxies the feedback energy mostly goes to the hot circumgalactic medium rather than directly impacts the dense and planar cold gas, thus making the feedback mainly preventive against further gas inflow.

\end{abstract}




\section{Introduction}\label{sec:intro}

The stellar kinematics and morphology encode the assembly history of galaxies.
In the consensus $\Lambda \mathrm{CDM}$ cosmology, galaxies build up by forming stars via accreted gas (\textit{in situ}) and by merging with other galaxies (\textit{ex situ}).
Because gas can cool efficiently, the \textit{in situ} star formation tends to establish dynamically cold disc structure where stars share the same sense of rotation.
By contrast, the violent relaxation during mergers blows stars further into the phase space (Section 4.10 of \citet{2008gady.book.....B}) and tends to form dynamically hot ellipsoidal/spheroidal structure like the bulges and halos.

The extensively accepted late- and early-type morphology classification in the Hubble scheme in principle differentiate galaxies with and without gas-related features in images such as spiral arms and extended dust lanes \citep{1961hag..book.....S}.
Although gas-rich spiral galaxies are generally disc-dominated, the late/early dichotomy itself does not guarantee a clean and clear classification of galaxy intrinsic morphology in terms of the disc-to-total mass ratio (D/T), and early-type S0 galaxies straddle almost the entire D/T range occupied by late-type disc galaxies \citep{2011MNRAS.416.1680C}.

In recent two decades, the integral field spectroscopy (IFS) has provided decisive constraints on assessing the intrinsic morphology of low-redshift galaxies \citep[see a review by][]{2016ARA&A..54..597C}.
Statistically significant datasets of spatially resolved stellar kinematics based on IFS reveal galaxy bimodality in stellar rotation.
While the slow rotator galaxy population show little to no rotation in their stellar velocity maps, the fast rotator population manifest clear hourglass-like velocity field, and the two populations are correspondingly quantified by low and high specific angular momenta.
This was shown by the first-generation IFS surveys, SAURON \citep{2002MNRAS.329..513D, 2007MNRAS.379..401E, 2007MNRAS.379..418C} and $\mathrm{ATLAS}^{\mathrm{3D}}$ \citep{2011MNRAS.413..813C, 2011MNRAS.414.2923K, 2011MNRAS.414..888E}, and then confirmed and further elaborated by subsequent IFS surveys MaNGA \citep{2015ApJ...798....7B, 2018MNRAS.477.4711G} and SAMI \citep{2015MNRAS.447.2857B, 2021MNRAS.505.3078V}.

A commonly used quantification of the level of stellar rotation is the normalized specific angular momentum, i.e. the spin parameter $\lambda_R$ \citep{2007MNRAS.379..401E}, which is by design close to the halo spin parameter \citep{1969ApJ...155..393P} in physical meaning.
Assessing galaxy intrinsic morphology from images alone bears strong degeneracy due to projection, and two thirds of visually classified elliptical galaxies show significant levels of stellar rotation in velocity maps \citep{2011MNRAS.414..888E}, indicating among them the presence of discs.
Additional kinematic information from IFS helps to break the degeneracy in the context of tensor virial theorem (TVT) that links together the level of rotational support, galaxy shape, and the anisotropy of velocity dispersion \citep{2005MNRAS.363..937B}.
On the $\lambda_R$ versus ellipticity $\varepsilon$ diagram, the population of fast rotators can be well described by TVT as oblate galaxies with radial anisotropy up to a maximum that depends on galaxy intrinsic shape $0.7 \times \varepsilon _{\mathrm{intr}}$ \citep{2007MNRAS.379..418C, 2021MNRAS.500L..27W}.
The slow-rotating ellipsoidal population occupy the low $\lambda_R$ and low $\varepsilon$ corner, and can be readily discriminated from relatively face-on discy galaxies given the large $\lambda_R$ difference at the same $\varepsilon$.
With the aid of IFS, the fast and slow classifications of galaxies are thus more precise in dividing galaxies of different intrinsic morphology.

The stellar spin measured within one half-light radius $\lambda_{R_{\rm e}}$ empirically has a primary dependence on stellar mass.
Slow rotators prevail in galaxy populations of stellar mass above a critical mass $\mathcal{M}_\mathrm{crit}\approx10^{11.3}\mathcal{M}_{\odot}$ \citep{2016ARA&A..54..597C, 2018MNRAS.477.4711G}.
Unlike slow rotators at lower mass, these massive slow rotators mostly host a stellar core that might be the imprint of major mergers \citep{2020A&A...635A.129K}.
Actively star-forming galaxies on the so-called star formation main sequence (SFMS) are generally spiral discy galaxies with high $\lambda_{R_{\rm e}}$.
At given mass, $\lambda_{R_{\rm e}}$ systematically decreases toward lower star formation levels \citep{2020MNRAS.495.1958W}.
This trend is sharp among massive galaxies but flat at low mass regime $\mathcal{M}_{\star}\sim10^{9.8}\mathcal{M}_{\odot}$ where discs are predominant even among low-mass passive galaxies whose star formation has been quenched.
Massive galaxies show weak but unambiguous environmental dependence of $\lambda_{R_{\rm e}}$, with lower values in denser regions \citep[e.g.,][]{2019arXiv191005139G, 2020MNRAS.495.1958W, 2021MNRAS.508.2307V}.
The environmental dependence for low-mass galaxies appears to be absent, under the low number statistics and the consequent large uncertainty.

A fundamental property of low-redshift galaxy populations is that their stellar spin is bimodal \citep{2016ARA&A..54..597C, 2018MNRAS.477.4711G, 2021MNRAS.505.3078V}.
We show \citep[][hereafter Wang23]{2023NAL000W} that the $\lambda_{R_{\rm e}}$ bimodality exists universally at all star formation levels, as well as in different environments.
The two kinematic subpopulations of the same mass and star formation level, with proved distinctness in the history of metal enrichment and recent star formation, suggest diverged evolutionary pathways for the quenching of star formation.
The metallicity of the slow-rotating population, as shown in Wang23, does not clearly depend on star formation level, in contrast to the fast-rotating galaxy population, and in particular passive slow-rotating galaxies are much less enriched than passive fast-rotating galaxies.
These observations point to the importance of gas outflows among the slow-rotating galaxies.

The full data release of MaNGA IFS survey \citep{2022ApJS..259...35A} and the third data release of its H\textsc{i} follow-up survey \citep[H\textsc{i}-MaNGA;][]{2019MNRAS.488.3396M, 2021MNRAS.503.1345S} have become available recently.
With the provided statistical power, it is now feasible to directly compare the cold gas content between fast- and slow-rotating populations under careful parameter control of mass, star formation, and environment.
In this work, we seek the impact of gas outflows among slow-rotating galaxies and show that they are generally more gas-poor than their fast-rotating counterparts.
We also compare the profile shape of the stacked spectra to coarsely infer the distribution and kinematics of H\textsc{i} gas.

Throughout this work we adopt the concordance cosmology with $H_0$ = $70\ \mathrm{km}\,\mathrm{s}^{-1}\,\mathrm{Mpc}^{-1}$, $\Omega_\mathrm{m}$ = 0.3, and $\Omega_{\Lambda}$ = 0.7.

\section{Sample}\label{sec:samp}

With the quality control introduced in \autoref{sec:samp} and \autoref{sec:meas}, our final sample is about $\sim4300$ galaxies that have all (1) robust $\lambda_{R_{\rm e}}$ measurements based on MaNGA kinematic maps, (2) good measurements of total stellar mass and star formation rate, and (3) robust and unambiguous H\textsc{i} single-dish spectra from H\textsc{i}-MaNGA survey.

\subsection{MaNGA integral field spectroscopy in the optical}\label{subsec:man}

To quantify the rotational support of galaxies, we take stellar kinematic maps from the complete MaNGA survey, which are publicly available as a part of the SDSS Data Release 17 \citep[DR17; ][]{2022ApJS..259...35A}.
MaNGA observed $\sim10000$ galaxies in the redshift range $0.01<z<0.15$ via the hexagonal integral field units with angular resolution FWHM of 2.5 arcsec and effective diameters ranging from 12 to 32 arcsec \citep{2015AJ....149...77D}.
Each spectrum covers 360-1030 nm with median instrument broadening $\sigma_{\mathrm{inst}}\,\sim\,72\,\mathrm{km}\,\mathrm{s}^{-1}$ \citep{2016AJ....152...83L} and typical spectral resolution $R \sim 2000$.

We choose galaxies in the MaNGA main sample: the Primary$+$ together with the Secondary sample.
Without cut to galaxy colour, morphology, or environment, galaxies in the two subsamples are mostly covered out to 1.5 and 2.5 half-light radii $R_{\rm e}$ and are selected to have a flat mass distribution.
See a detailed target selection scheme in \citet{2017AJ....154...86W}.
After discarding duplicates and galaxies with problematic redshifts, we get a sample of 9793 galaxies with kinematic maps recorded in the MaNGA MAPS files.

We cross match the MaNGA sample with the version X2 of GALEX-SDSS-WISE Legacy Catalogue \citep[GSWLC-X2,][]{2016ApJS..227....2S,2018ApJ...859...11S} to get total stellar mass and total star formation rate (SFR) for MaNGA galaxies, as MaNGA IFS alone still misses a significant fraction of light in the outer part of galaxies.
Briefly, the stellar mass and SFR are derived by fitting the spectral energy distribution of ultraviolet, optical, and mid-infrared fluxes, using one of the state-of-the-art modelling tools CIGALE \citep{2009A&A...507.1793N, 2019A&A...622A.103B}.

The star formation state of galaxies is quantified by the distance from the SFMS (see more details in Wang23) at given stellar mass $\Delta_\mathrm{MS}$:
\begin{align}
& \Delta_\mathrm{MS}\equiv \mathrm{lg}\,\mathrm{SFR}-\mathrm{lg}\,\mathrm{SFR}_\mathrm{MS}(\mathcal{M}_{\star})\ \mathrm{,} \\[4pt]
& \mathrm{lg}\,\mathrm{SFR}_\mathrm{MS}=\begin{cases}
    0.8285\times\mathrm{lg}\, \mathcal{M}_{\star}- 8.2963, & \mathrm{lg}\, \mathcal{M}_{\star}<10\mathrm{,}\\
    0.3144\times\mathrm{lg}\, \mathcal{M}_{\star} - 3.1687, & \mathrm{lg}\, \mathcal{M}_{\star}\geq 10\mathrm{.}
  \end{cases}
\end{align}
Stellar mass and star formation rate are respectively in unit of solar mass and solar mass per year.
We define those with $-0.35<\Delta_\mathrm{MS}<0.35$, i.e. within the typical scatter of the SFMS, as normal star-forming galaxies.
Galaxies with SFR less than ten percent of star-forming galaxies on the SFMS, $\Delta_\mathrm{MS}<-1$, are considered passive galaxies.
Other galaxies in between with $-1<\Delta_\mathrm{MS}<-0.35$ are defined as green valley galaxies.

\subsection{H\textsc{i}-MaNGA single-dish spectroscopy in the L-band}\label{subsec:himan}

H\textsc{i}-MaNGA is an atomic hydrogen 21cm line follow-up survey for MaNGA, using the single-dish Robert C. Bryd Green Bank Telescope (GBT).
Essentially, all MaNGA galaxies out of the footprint of the large blind H\textsc{i} survey ALFALFA \citep{2011AJ....142..170H} and with redshifts $z<0.05$ are selected for observation.
Therefore, the sample are desirably not flux limited but miss the many massive MaNGA galaxies at higher redshifts.
The GBT observations are designed to reach rms noises $\sim1.5\,\mathrm{mJy}$ at $10\,\mathrm{km}\,\mathrm{s}^{-1}$ resolution, comparable to the noises of the ALFALFA survey, so that the existing ALFALFA data and the newly acquired data from GBT observations form a homogeneous dataset.

The third data release of H\textsc{i}-MaNGA comprises 6632 H\textsc{i} observations for MaNGA galaxies, 3358 of which are conducted by GBT.
9\% are excluded in our analyses because they have either complex baseline in the spectra, or underestimated flux due to detections in the OFF observation, or are contaminated seriously that more than 20\% of the total flux comes from companions.

\section{Measurements}\label{sec:meas}

\subsection{Stellar spin $\lambda_{R_{\rm e}}$ for $9793$ MaNGA galaxies}\label{subsec:lam}

\begin{table*}
\caption{Table of photometric and kinematic properties for the 9793 MaNGA galaxies in our main sample.
Column (1): The ID associated with a specific observation for a unique MaNGA galaxy.
Column (2), Column (3) and Column (4) are respectively right ascension and declination in J2000, and redshift of the galaxy center.
Column (5) and Column (6) list the photometric and kinematic position angle defined as East of North, determined respectively via \textsc{MgeFit} and \textsc{PaFit} \citep{2006MNRAS.366..787K} available at https://pypi.org/user/micappe/.
Column (7) and Column (8) are the semimajor axis of and the ellipticity in the MGE half-light isophote (see \autoref{subsec:lam}).
Column (9) lists the stellar spin parameter corrected for beam smearing, measured within the half-light ellipse and Column (10) gives the bad pixel fraction within the half-light ellipse (flagged by the MaNGA DAP as DONOTUSE).
Column (11) lists the FWHM of the MaNGA PSF.
Column (12) is the 5 digits bitmask recorded as a decimal number (e.g., a recorded decimal number 2 points to the bitmask 00010). Every one of the five bits mask if ($0/1\rightarrow \mathrm{False}/\mathrm{True}$) the galaxy is in a certain situation (bit/bit/bit/bit/bit$\rightarrow$ situation (v)/(iv)/(iii)/(ii)/(i)) described in the quality control (see \autoref{subsec:lam}).
A simple way of converting the decimal number into the 5 digits bitmask is to use \textsc{NumPy} function \textsc{binary\_repr} vectorized by \textsc{NumPy} function \textsc{vectorize}.
Column (13) flags whether the $\lambda_{R_{\rm e}}$ measurements are robust according to the quality control.
Column (14) lists the visual kinematic classifications based on stellar velocity and velocity dispersion maps.
"R", regular and hourglass-like rotation;
"NR", no clear rotation;
"KDC", kinematic distinct core that shows central rotation detached from outer kinematics;
"PR", polar rotation, perpendicular inner and outer rotation;
"CR", counterrotation in the velocity field;
"2sigma", two-sigma galaxies \citep{2011MNRAS.414.2923K} that exhibit two symmetric peaks in velocity dispersion maps, indicating the presence of counterrotating discs;
"unclassifiable", kinematics that does not fit in the above classifications.
A complete table is available from the journal website.}
\begin{center}
\begin{tabular}{c c c c c c c c c c c c c c}
 \hline
Plate-IFU & RA & DEC & $z$ & $\Psi_\mathrm{phot}$ & $\Psi_\mathrm{kine}$ & $R_\mathrm{e}^\mathrm{maj}$ & $\varepsilon$ & $\lambda_{R_{\rm e}}$ & $f_\mathrm{dnu}$ & PSF$_\mathrm{FWHM}$ & Bitmask & Clean & Kinematic\\
& $[^{\circ}]$ & $[^{\circ}]$ &  & $[^{\circ}]$ & $[^{\circ}]$ & [arcsec] & & & & [arcsec] &  & sample? & classification\\
(1) & (2) & (3) & (4) & (5) & (6) & (7) & (8) & (9) & (10) & (11) & (12) & (13) & (14)\\ \hline
8131-3703 & 112.1694 & 41.4730 & 0.1191 & 82.87 & 86.97 & 3.97 & 0.146 & 0.194 & 0 & 2.44 & 0 & y & R\\
8131-12702 & 112.3603 & 40.1692 & 0.0735 & 12.85 & 52.58 & 5.44 & 0.045 & 0.138 & 0.011 & 2.38 & 1 & n & unclassifiable\\
8131-6102 & 112.0707 & 40.0841 & 0.0495 & 75.69 & 66.74 & 6.31 & 0.173 & 0.174 & 0 & 2.39 & 0 & y & R\\
8131-6103 & 112.6707 & 40.1918 & 0.1201 & 157.46 & 38.43 & 6.18 & 0.146 & 0.125 & 0 & 2.41 & 0 & y & R\\
8131-12703 & 111.6757 & 41.2235 & 0.0431 & 88.48 & 93.03 & 7.58 & 0.707 & 0.847 & 0 & 2.44 & 0 & y & R\\
8131-3702 & 111.5165 & 40.5580 & 0.0502 & 76.80 & 88.99 & 3.19 & 0.277 & 0.468 & 0 & 2.34 & 0 & y & R\\
8131-3704 & 111.9753 & 40.0920 & 0.1209 & 90.61 & 99.10 & 4.29 & 0.050 & 0.110 & 0.023 & 2.45 & 0 & y & NR\\
8131-12705 & 112.5599 & 39.5335 & 0.1415 & 86.07 & 117.30 & 3.31 & 0.034 & 0.043 & 0 & 2.42 & 0 & y & NR\\
8131-12704 & 112.4677 & 39.8915 & 0.0498 & 1.90 & 2.02 & 9.40 & 0.673 & 0.89 & 0 & 2.41 & 0 & y & R\\
8131-6101 & 111.0908 & 39.0190 & 0.0519 & 124.59 & 129.44 & 3.79 & 0.437 & 0.813 & 0 & 2.45 & 0 & y & R\\
8131-3701 & 111.8748 & 38.6996 & 0.0807 & 55.84 & 54.61 & 7.07 & 0.337 & 0.855 & 0.123 & 2.49 & 0 & y & R\\
8083-6102 & 51.1150 & -0.0863 & 0.0365 & 65.86 & 66.74 & 6.43 & 0.464 & 0.733 & 0 & 2.43 & 0 & y & R\\
\hline
\end{tabular}
\end{center}
\label{tab1}
\end{table*}

Mostly following the procedures in \citet{2018MNRAS.477.4711G} for MaNGA galaxies in SDSS DR14 \citep{2018ApJS..235...42A}, we measure $\lambda_{R_{\rm e}}$ for the 9793 MaNGA galaxies in DR17.
We take the MAPS files that catalogue the maps of stellar kinematics, which are produced by the Data Analysis Pipeline of MaNGA \citep[DAP; ][]{2019AJ....158..231W}.
With the Voronoi spatial binning method \citep{2003MNRAS.342..345C}, DAP divides the data cubes to achieve a minimum signal-to-noise ratio of $\sim10$ per spectral bin of width $70\,\mathrm{km}\,\mathrm{s}^{-1}$.
Then DAP fits a set of 49 families of stellar spectra from the MILES stellar library \citep{2006MNRAS.371..703S, 2011A&A...532A..95F}, convolved with a Gaussian function for the line-of-sight velocity distribution (LOSVD) of stars, to the observed absorption-line spectra via the Penalised Pixel-Fitting method \citep{2004PASP..116..138C, 2017MNRAS.466..798C}.
The stellar kinematics at each pixel is characterized by the mean stellar velocity $V$ and velocity dispersion $\sigma$ of the derived LOSVD.
We correct the systemic velocity by bisymmetrizing the velocity field using routine\footnote{Available at https://pypi.org/project/pafit/} \textsc{fit\textunderscore kinematic\textunderscore pa} \citep{2006MNRAS.366..787K}, and correct velocity dispersion for instrument broadening by the values in entry \texttt{STELLAR\textunderscore SIGMACORR} of MaNGA MAPS files.
For galaxies with intrinsically low velocity dispersion (low-mass or disc-dominated), the estimated corrections for dispersion can be larger than the raw measurements which results in negative values of intrinsic dispersion.
We use zero values instead for these pixels.
The spin parameter $\lambda_{R_{\rm e}}$ is calculated using equation 5 and 6 of \citet{2007MNRAS.379..401E}:
\begin{equation}
\lambda_{R}\,\equiv \, \frac{\left< R\,|V| \right>}{\left< R\, \sqrt{V^2+\sigma ^2} \right>}\,=\,\frac{\sum_{n=1}^{N} F_n R_n |V_n|}{\sum_{n=1}^{N} F_n R_n \sqrt{V_n^2+\sigma _n ^2}}
\end{equation}
where $F_n$, $V_n$, and $\sigma _n$ are the flux, mean velocity and velocity dispersion of the nth pixel.
The summation is performed over the $N$ pixels within the elliptical half-light radius $R_{\rm e}$.

We take the half-light ellipses determined for MaNGA DR17 from \citet{2023MNRAS.522.6326Z}.
Briefly, the SDSS $r$-band photometry from NASA-Sloan Atlas\footnote{http://www.nsatlas.org}
\citep[NSA; ][]{2011AJ....142...31B} is fitted with the Multi-Gaussian Expansion (MGE) method\footnote{Using the \textsc{MgeFit} Python software package of \citet{2002MNRAS.333..400C} available at https://pypi.org/project/mgefit/} \citep{1994A&A...285..723E, 2002MNRAS.333..400C}.
The isophote containing half of the MGE total luminosity is determined by implementing the steps (i) to (iv) described before equation 12 of \citet{2013MNRAS.432.1709C}.
The ellipticity $\varepsilon$ is then calculated inside the half-light isophote as the second moments of light distribution:
\begin{equation}
(1-\varepsilon)^2\,=\,q^{\prime 2}\,=\,\frac{\left< y^2 \right>}{\left< x^2 \right>}\,=\,\frac{\sum_{k=1}^{P} F_k y_k^2}{\sum_{k=1}^{P} F_k x_k^2}
\end{equation}
where $F_k$ and $(x_k,y_k)$ are the flux and coordinates of the $k$th pixel.
$\varepsilon$ and the area of the half-light isophote define the half-light ellipse within which $\lambda_{R_{\rm e}}$ is calculated.

We apply quality control for $\lambda_{R_{\rm e}}$ measurements by excluding galaxies in the following situations:
\begin{enumerate}[(i)]
 \item Kinematics is contaminated by close companion galaxies.

 \item MGE fit to the photometry is obviously incorrect from visual inspection, mostly due to surrounding objects.

 \item Kinematic maps are problematic due to, e.g., low surface brightness that results in just one or few Voronoi bins or the presence of active galactic nuclei that makes fatal fits to the spectra.

 \item Ellipticity $\varepsilon>0.4$ and the misalignment between the kinematic and photometric major axes is greater than 30 degrees. See Appendix A of \citet{2018MNRAS.477.4711G}.

 \item More than 30\% of pixels within the half-light ellipse have $\sigma_0<\sigma_\mathrm{corr}$, where $\sigma_0$ is the raw measurement of velocity dispersion and $\sigma_\mathrm{corr}$ is the correction for instrument broadening.
 We do not exclude visually identified disc-dominated galaxies that meet this criterion, given their intrinsically small velocity dispersion relative to rotation. See Appendix B of \citet{2018MNRAS.477.4711G}.

 \item More than 20\% of pixels within the half-light ellipse are flagged by the MaNGA DAP as DONOTUSE (i.e. the entry \texttt{MANGA\textunderscore DAPPIXMASK} value no less than 1073741824).

 \item Galaxies are too small that $R_\mathrm{e}^\mathrm{maj}/\sigma_\mathrm{PSF}<3/2$, where $R_\mathrm{e}^\mathrm{maj}$ is the major axis of the MGE half-light isophote and $\sigma_\mathrm{PSF}$ is the size of MaNGA synthesized beam (i.e. $0.425\times$ FWHM of the PSF).
\end{enumerate}
Such quality control eliminates galaxies with apparently questionable $\lambda_{R_{\rm e}}$ measurements which make up 12\% of the 9793 MaNGA galaxies.
Situations (iii) and (v) are more frequent among low-mass galaxies, and we limit our analyses above a stellar mass $\mathcal{M}_{\star}=10^{9.5}\mathcal{M}_{\odot}$.
Adjusting individual threshold in the quality control does not affect the conclusion of this work.

The smearing of rotation field due to finite spatial resolution is corrected for using the analytic functions derived in \citet{2018MNRAS.477.4711G}, which has been tested in simulations \citep{2019MNRAS.483..249H}.
The correction is only applied to galaxies which show regular, hourglass-like rotation (see the panel e of Figure 4 in \citet{2016ARA&A..54..597C}) in their velocity field.
Nonregular rotator galaxies are mostly slow-rotating with low $\lambda_{R_{\rm e}}$ and their underestimation of $\lambda_{R_{\rm e}}$ due to smearing is small \citep{2020MNRAS.497.2018H} and generally negligible.

\begin{figure}
 \includegraphics[width=0.45\textwidth]{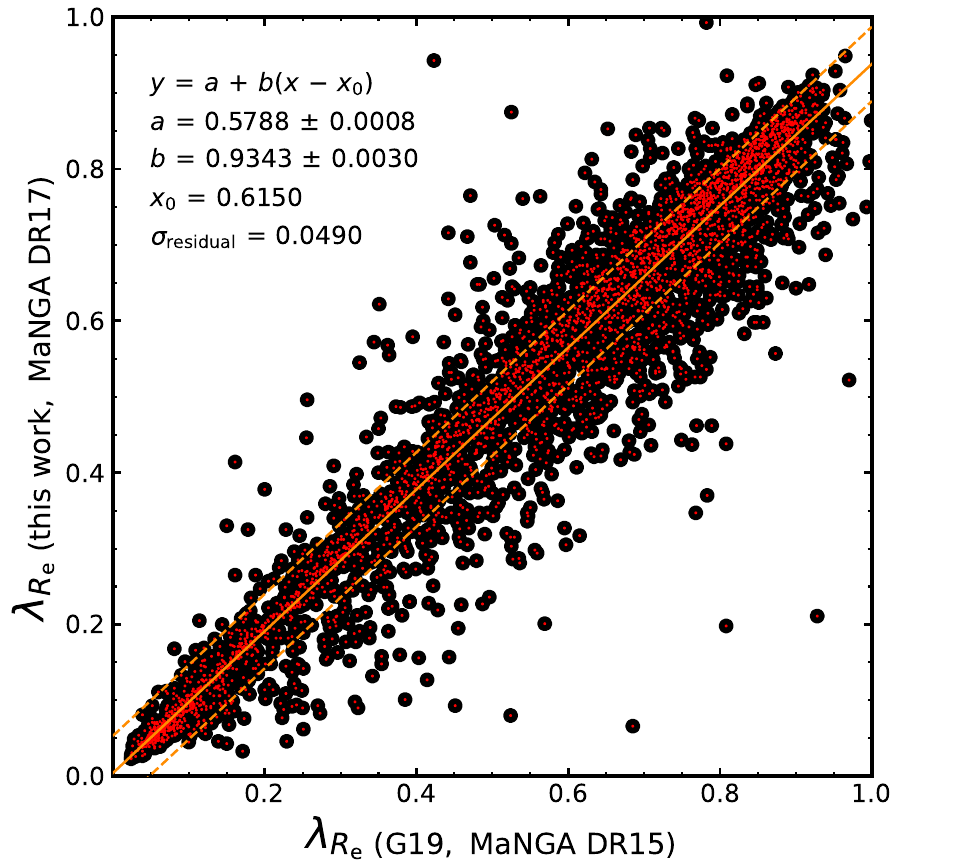}
 \caption{A comparison of seeing corrected $\lambda_{R_{\rm e}}$ between \citet{2019arXiv191005139G} for MaNGA DR15 and this work for MaNGA DR17, excluding galaxies by the quality control described in the text.
 Each galaxy is represented by a red dot with thick black edge.
 A robust linear fit to the relation and the standard deviation of the residuals are shown by the orange solid line and dashed lines respectively.
 The best fitting parameters are shown in the upper left corner.}
 \label{fig:comp}
\end{figure}

In \autoref{fig:comp} we show an one-to-one comparison of $\lambda_{R_{\rm e}}$ measurements between \citet{2019arXiv191005139G} for MaNGA DR15\footnote{The catalogue can be found in \citet{2022MNRAS.511..139B}.} (following \citet{2018MNRAS.477.4711G} for DR14) and this work for DR17.
We fit the relation using the robust line fitting algorithm \textsc{LtsFit}\footnote{https://pypi.org/project/ltsfit/} \citep[Section 3.2 of][]{2013MNRAS.432.1709C}.
The general consistency is good, with small residuals of scatter $\sim0.05$ from the nearly linear relation.
There are about 10 galaxies apparently above $5\sigma$ of the relation.
Many of them are low-mass galaxies of low surface brightness, bearing large uncertainty in getting their intrinsic velocity dispersion, and can be excluded by the low mass limit for our analyses.

\subsection{Mean gas fractions via stacking technique}\label{subsec:stac}

To thoroughly compare the H\textsc{i} gas reservoir of fast- and slow-rotating galaxy populations, we take the non-detections into account via the stacking technique, which effectively elevates signal-to-noise ratio as noises cancel out due to their nearly stochastic nature.

We follow \citet{2011MNRAS.411..993F} to measure the average gas mass fractions.
Specifically, for galaxies in certain bins of stellar mass, star formation level, and stellar kinematics, we first shift each H\textsc{i} spectrum to rest frequency according to the optical redshift of the galaxy center.
Then we stack the spectra of gas mass fraction weighted by the noise level rms:
\begin{align}
& S_\mathrm{stack}\,=\,\frac{\sum_{i=1}^{N} w_i S_i^{\prime}}{\sum_{i=1}^{N} w_i}\ \mathrm{,} \\[4pt]
& S_i^{\prime}\,=\,\frac{S_i D_\mathrm{L}^2(z_i)}{\mathcal{M}_{\star,i}}\left(\mathrm{Jy}\,\mathrm{Mpc}^2\,\mathcal{M}_{\odot}^{-1}\right)\ \ \ \ \mathrm{and}\ \ \ \ w_i\,=\,\frac{1}{\mathrm{rms}^2_i},
\end{align}
where $D_\mathrm{L}$ is the luminosity distance at redshift $z_i$, and $\mathcal{M}_{\star,i}$ is the stellar mass of galaxy $i$.
From the stacked spectra, we visually define a H\textsc{i} line detection (or non-detection).
For detections, the emission line is integrated and multiplied by a factor of $2.356\times10^5$ (equation 3 of \citet{2011MNRAS.411..993F}) to finally convert to H\textsc{i} mass to stellar mass ratio.
The uncertainty of the derived mean gas mass fraction is assessed via bootstrapping, i.e. by resampling the spectra in each stack.
While for non-detections, we assume a conservative upper limit as a $5\sigma$ signal with a width of 300 km s$^{-1}$ (see equation 9 of \citet{2019MNRAS.488.3396M}).

\begin{figure*}
  \begin{center}
 \includegraphics[width=0.75\textwidth]{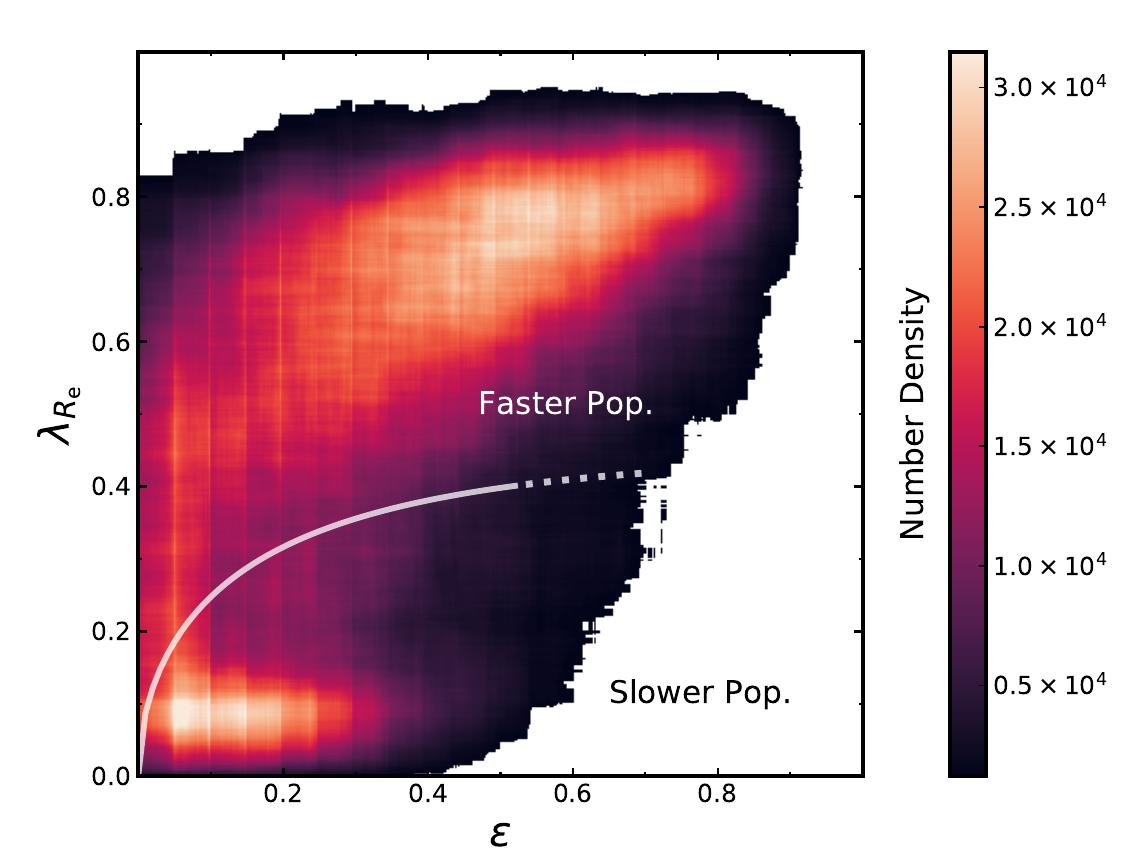}
 \caption{The galaxy number surface density on the $(\lambda_{R_{\rm e}},\varepsilon$) diagram, based on full MaNGA excluding galaxies by the quality control described in \autoref{subsec:lam}.
 The density is calculated in a box of width 0.1 at the step of 0.002.
 The white solid curve is the TVT predicted locus for an oblate galaxy with $\lambda_{R,\mathrm{intr}}=0.4$ and $\varepsilon_\mathrm{intr}=0.525$ at anisotropy $\delta=0.7\varepsilon_\mathrm{intr}=0.3675$, projected at all inclinations, and the white dotted line is a crude linear extension.
 Throughout this work, we define the galaxies above/below the white curve as faster/slower population.
 }
 \label{fig:le}
\end{center}
\end{figure*}

\section{Results}\label{sec:resu}

Wang23 presents the ubiquitously bimodal $\lambda_{R_{\rm e}}$ distribution of galaxy populations across a large range of star formation level and environment\footnote{The $\lambda_{R_{\rm e}}$ bimodality exists in diverse environment as, for example, quantified by group richness, dark matter halo mass, and galaxy number density.}, and shows that generally an intrinsic stellar bin $\lambda_{R_{\rm intr}}=0.4$ empirically separates galaxies of fast and slow rotation.
In \autoref{fig:le}, we illustrate this empirical result using the full MaNGA.

\begin{figure*}
 \includegraphics[width=0.95\textwidth]{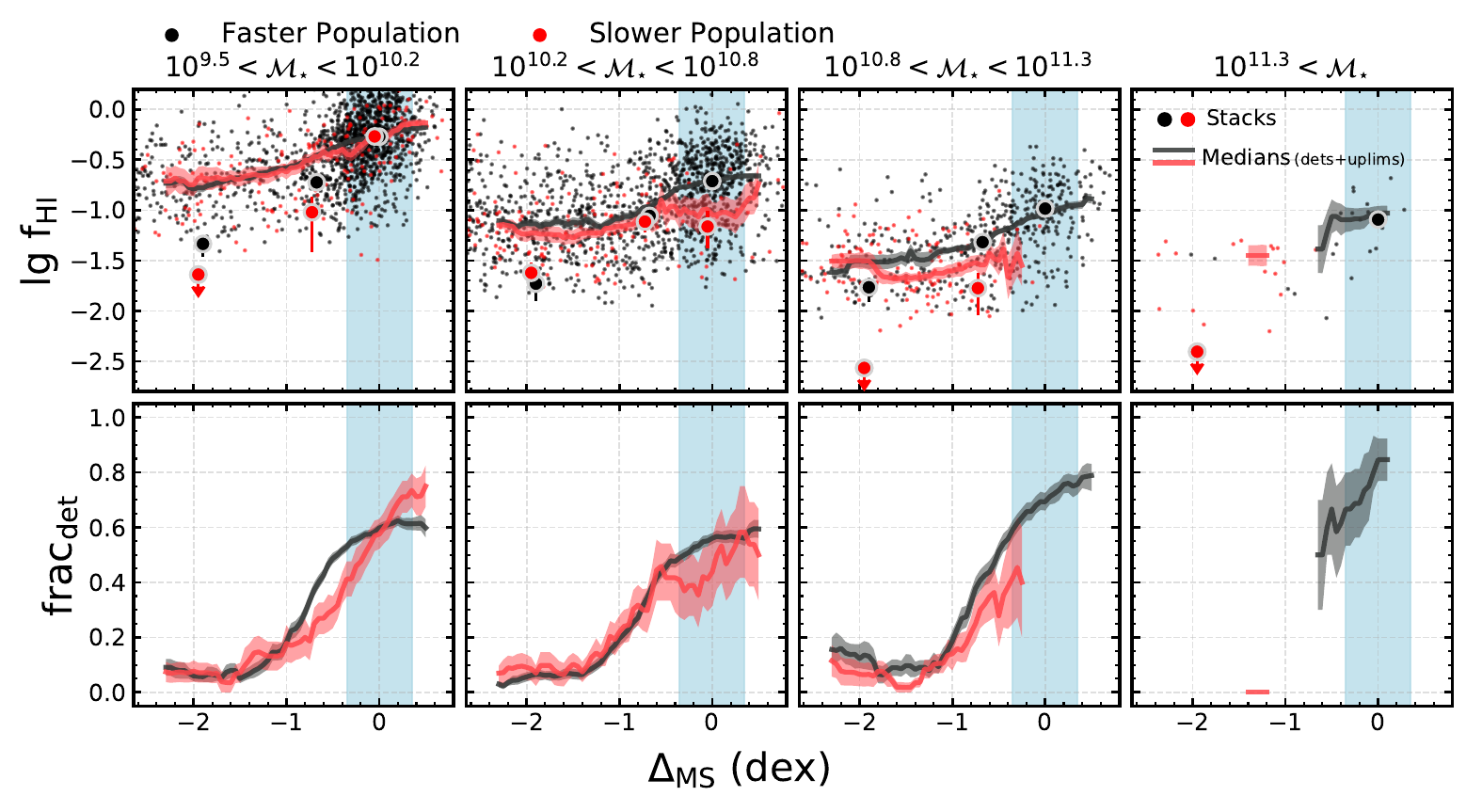}
 \caption{The dependence of H\textsc{i} gas reservoir on the stellar rotation of galaxies.
 Upper row: H\textsc{i} to stellar mass ratio on logarithmic scale as a function of star formation level $\Delta_\mathrm{MS}$, for the galaxy population of faster (black) and slower (red) rotation respectively.
 Each column shows the relation in a certain stellar mass bin, with stellar mass increasing rightward.
 Small dots are for individual galaxies and include both detections and upper limits, while the solid lines and shaded envelopes show the running medians and uncertainties (standard deviation of 1000 bootstrap samples) in sequential $\Delta_\mathrm{MS}$ windows of width 0.8 dex at steps of 0.05 dex and of at least 10 galaxies.
 Large points show the mean gas fractions with uncertainties estimated via the standard stacking technique in the three broad $\Delta_\mathrm{MS}$ bins for star-forming (i.e. on the SFMS, the blue band), green valley, and passive galaxies defined in \autoref{subsec:man}.
 The bins of galaxies fewer than 10 are not shown.
 Lower row: The H\textsc{i} detection ratio as a function of $\Delta_\mathrm{MS}$ for faster (black) and slower (red) population with estimated uncertainty as the inner 68\% of 1000 bootstrap samples.
 }
 \label{fig:fhi}
\end{figure*}

\begin{figure}
 \includegraphics[width=0.4\textwidth]{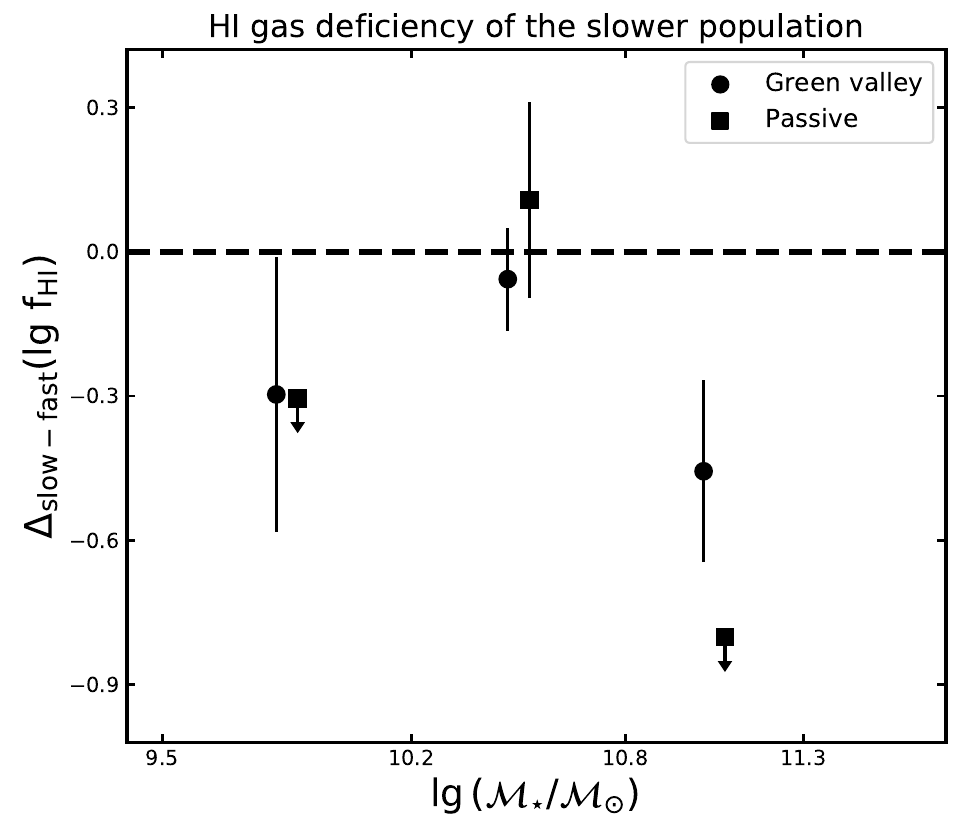}
 \caption{The H\textsc{i} gas deficit of the slower population, i.e. lg$\frac{f_\mathrm{H\textsc{i},slow}}{f_\mathrm{H\textsc{i},fast}}$ based on the stacking derived quantities in \autoref{fig:fhi}, as a function of stellar mass, for green valley (points) and passive (squares) galaxies respectively.
 The error bars show propogated errors as the root of quadratic sum of the errors for faster and slower populations respectively.
 }
 \label{fig:dfhi}
\end{figure}

\autoref{fig:le} shows galaxy number density on the $(\lambda_{R_{\rm e}},\varepsilon$) diagram and it is strikingly apparent that galaxies in the local Universe evolve in two distinct forms as can be well distinguished as two kinematic populations on the diagram.
While the slow-rotating galaxies crowd in the lower left corner of the $(\lambda_{R_{\rm e}},\varepsilon$) diagram, the fast-rotating galaxies mainly populate the upper right area with their relatively face-on peers extending as a tail toward the slow-rotating population.
The area in between is loosely populated by galaxies of comparable disc and spheroid components.

The white curve shows the TVT predicted locus for a axisymmetric galaxy with $\lambda_{R,\mathrm{intr}}=0.4$ (at anisotropy $\delta=0.7 \times \varepsilon _{\mathrm{intr}}$) at all inclinations, which indeed broadly separates the two kinematic populations.
We use this curve\footnote{The \textsc{python} function used to generate the curve can be found at https://github.com/Santiago-WANG/Divider} hereafter to respectively define the galaxy population of faster (above the curve) and slower (below) rotation, abbreviated as faster and slower population.
The naming is intentionally different from the classic fast and slow rotators \citep{2016ARA&A..54..597C}.
The classic definition of slow rotators seeks to capture galaxies without stellar disc, i.e. those dominated entirely by hot spheroid structure.
Here we base the definition more on the empirical bimodality which classifies galaxies with tiny but nonzero D/T also into the slower population.
Given the small number of galaxies in the intermediate population, such minor difference in fast/slow definition does not affect our conclusion.

\begin{figure*}
 \includegraphics[width=0.95\textwidth]{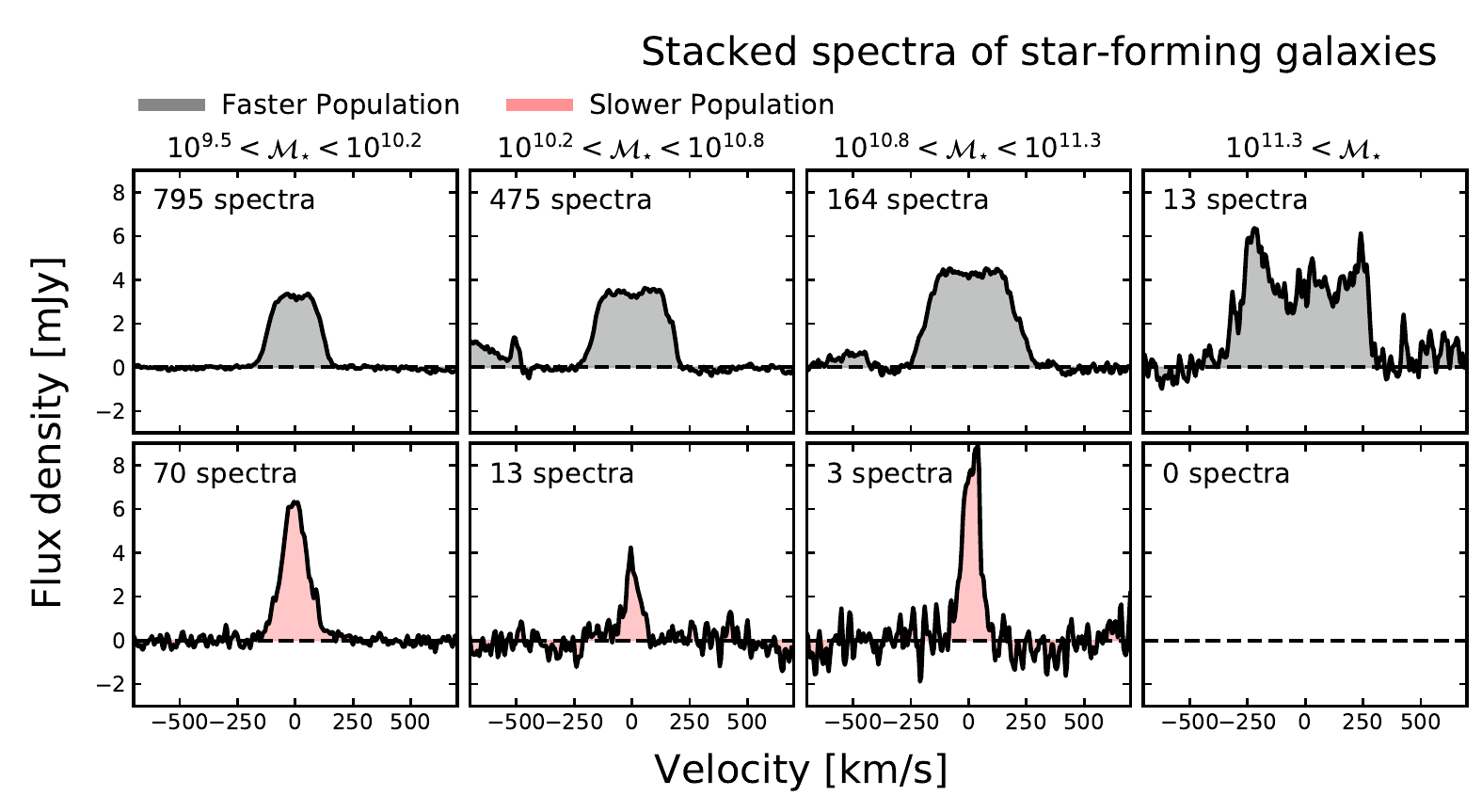}
 \caption{Spectra raw (i.e. without weighting) stacks of star-forming faster (upper row) and slower (lower row) populations.
 The number of galaxies in the stack is denoted in each panel.
 }
 \label{fig:prof}
\end{figure*}

In \autoref{fig:fhi} we turn to the main goal of this work and study the dependence of H\textsc{i} gas reservoir on the stellar rotation of galaxies.
The upper and lower rows show the H\textsc{i} to stellar mass ratio and the fraction of galaxies with detected H\textsc{i} as a function of $\Delta_\mathrm{MS}$, with each column for the relation in a certain stellar mass bin, and black and red colour respectivevly for the galaxy population of faster and slower rotation.
In each panel of the upper row, the small dots are for individual galaxies based on the H\textsc{i} gas mass (including both detections and upper limits) catalogued in H\textsc{i}-MaNGA, while the solid lines and shaded envelopes show the running medians and associated uncertainties in sequential $\Delta_\mathrm{MS}$ windows of width 0.8 dex at steps of 0.05 dex and of at least 10 galaxies.
In the three broad $\Delta_\mathrm{MS}$ bins for star-forming, green valley, and passive galaxies defined in \autoref{subsec:man}, we show the mean gas fractions estimated via the stacking technique as large points (not for bins of galaxies fewer than 10), which take proper account of the non-detections.

The incompatible redshift ranges at high stellar mass end of MaNGA and H\textsc{i}-MaNGA, together with the fact that the slower population comprise a small fraction on the SFMS (shown by the blue band), make unattainable a statistically significant conclusion about the kinematic dependence of H\textsc{i} as a function of mass for star-forming galaxies.
There are more than 60 slow-rotating galaxies on the SFMS in the lowest mass bin, and all indicators (i.e. the medians, detection fractions, and stacks) consistently show that low-mass star-forming galaxies of faster and slower rotation have the same H\textsc{i} gas abundance.
At higher mass $10^{10.2}<\mathcal{M}_{\star}/\mathcal{M}_{\odot}<10^{10.8}$, the star-forming galaxies of slower rotation are critically more gas-poor than their fast-rotating counterparts, with large uncertainty due to the small number of slow-rotating galaxies.
We refrain from drawing a conclusion on the SFMS, and plan to enrich our sample at high mass end using observations by The Five-hundred-meter Aperture Spherical radio Telescope \citep{2019SCPMA..6259502J}.

We thus focus on galaxy populations below the SFMS, where the samples of both faster and slower populations are large enough (except for the highest mass bin) for statistically robust conclusion.
Below the SFMS, the frequency of H\textsc{i} detections drops sharply, down to below 20\% among passive galaxies, so that the medians deviate from the values estimated by stacking and are no longer fully informative about the locations of the real distributions.
Taking non-detections into proper account, the stacks exhibit a clear mass trend that is coherent among both green valley and passive galaxies.
In the lowest mass bin, the slower population below the SFMS have less H\textsc{i} than the faster population at the same star formation level.
This cold gas deficit is more significant for massive galaxies in the mass range $10^{10.8}<\mathcal{M}_{\star}/\mathcal{M}_{\odot}<10^{11.3}$, which also manifests in the medians and detection fractions even in such low detection fraction regime.
In these two mass bins, the reduction of gas fraction with decreasing SFR appears to be sharper in slower population than in faster population.
However, in the intermediate mass range $10^{10.2}<\mathcal{M}_{\star}/\mathcal{M}_{\odot}<10^{10.8}$ the H\textsc{i} gas reservoir is indistinguishable between galaxy populations of faster and slower rotation.
These results and the following remain qualitatively the same when we study central galaxies only, using \citet{2007ApJ...671..153Y} group catalogue, to exclude potentially overwhelming environmental influence on the satellites.
While low-mass central galaxies are mostly in isolation, massive centrals can either host small groups or rich clusters.
We thus further restrict the analysis for massive galaxies in narrow dark matter halo mass bin, and we find that the stacking based gas fractions vary only at the level of $\sim0.1$ dex.
This is not unexpected given the weak environmental dependence of stellar kinematics at given mass and SFR \citep{2020MNRAS.495.1958W}.

The H\textsc{i} gas deficit of the slower population, as logarithmic difference of stacking derived mean gas fractions from the faster counterparts, is highlighted as a function of stellar mass in \autoref{fig:dfhi}.
The mass trend is in n shape, where the gas deficit disappears in the intermediate mass range for both green valley and passive galaxies.
At lower mass, the H\textsc{i} reservoir of slow-rotating galaxies is only half the amount of that of fast-rotating galaxies given the same star formation, though under large uncertainty among the low-mass green valley galaxies.
The gas deficit is more severe for massive galaxies, and particularly, the H\textsc{i} gas mass of slow-rotating galaxies of stellar mass $\sim10^{11}\mathcal{M}_{\odot}$ is almost reaching an order of magnitude less than fast-rotating galaxies.

The H\textsc{i} gas deficit provides further observational evidence for an important role that outflows paly in the star formation quenching of slow-rotating galaxies.
It is very likely that slow-rotating galaxies, compared to their fast-rotating peers, experience more episodes of violent evolution when they are far away from dynamical equilibrium, during which the H\textsc{i} gas is disturbed to a hotter state and may be spatially well coupled with baryonic feedback processes.
This is a practical scenario for the cold gas deficiency of the slower population, and is partly testable using the H\textsc{i} spectra from H\textsc{i}-MaNGA.
Even though without spatial resolution, the spectra record the velocity distribution of H\textsc{i} gas and can be used to infer the kinematic state and spatial distribution considering the general shape of galaxy circular velocity curves.
A thick component of cold gas lags behind the planar gas in rotational velocity \citep{2006MNRAS.366..449F}.
Gas in more concentrated configurations can have lower velocity as the circular velocity curve is still rising in the inner several kpc.
As a consequence, concentrated thick gas which couples well with, for example, biconical wind from active galactic nuclei, is more likely to show single-peaked velocity distributions.

\begin{figure*}
 \includegraphics[width=0.95\textwidth]{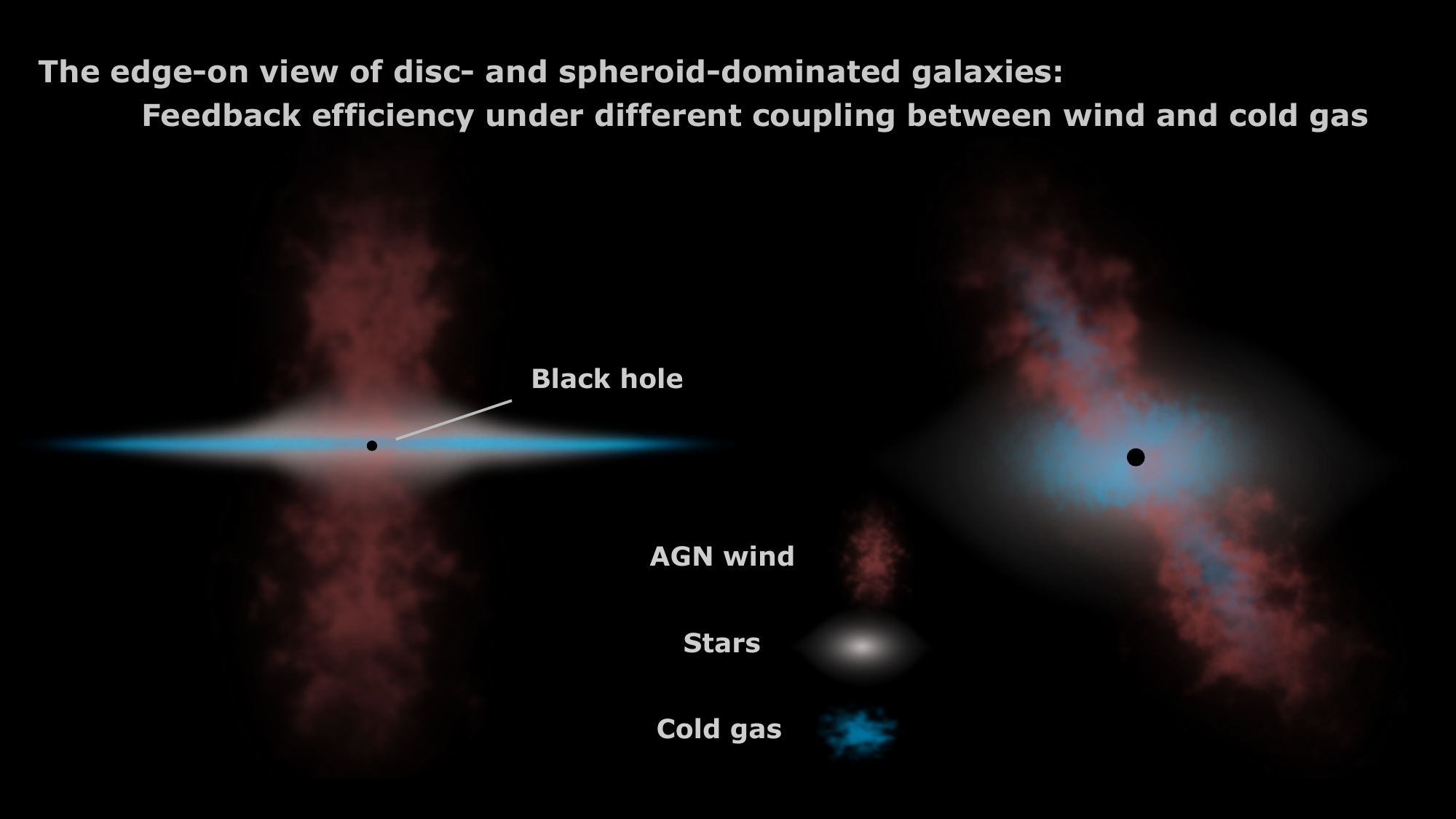}
 \caption{An illustration of how disturbed cold gas away from a planar form can elevate feedback efficiency.
 This figure shows, from an edge-on perspective, a disc-dominated (left-hand side) and a spheroid-dominated (right-hand side) galaxy with four components including a central black hole (black blob), stellar body (silk), cold gas (blue), and plasma wind from an active galactic nucleus (red).
 A disc galaxy without much gravitational interference has its cold gas settled in a planar form over most of the evolution history.
 The biconical wind launched by the central black hole then has little surface of impact to drive massive gas outflows.
 While a spheroidal galaxy, which must have experienced significant mergers over time, tends to possess volume filling gas in hotter kinematics and therefore its gas couples well with the wind.
 Moreover, a more massive bulge/spheroid means a more massive black hole with higher energy output.
 Jet is not shown here but would have the same effect.
 The wind/jet direction tends to be aligned more with the spin axis of fast-rotating galaxies than with slow-rotating galaxies that have little sense of bulk rotation \citep{2022arXiv221113614B}.
 }
 \label{fig:draw}
\end{figure*}

This is indeed what we found among slow-rotating galaxies, but for star-forming ones only.
\autoref{fig:prof} shows the stacked spectra of star-forming faster (upper row) and slower population (lower row) in the four mass bins separately.
The stacking here applies no weighting as we care about the spectrum shapes rather than luminosity, but the original weighting scheme gives the same conclusion.
The shape difference between the two kinematic populations is apparent.
The stacked spectra of fast-rotating galaxies are broad and all show two maxima on the sides, keeping the main feature of relatively edge-on gas discs (i.e. the double horns).
By contrast, star-forming slow-rotating galaxies universally exhibit narrow and single-peaked spectra.
The small sample size in three of the bins, especially for the massive slow-rotating galaxies, potentially makes some of the profile shapes biased toward certain inclinations.
However, the adopted kinematic classifications are inclination insensitive, and the consistency across all stellar mass ranges suggests that this general difference in profile shape is robust against projection biases.

\section{Discussion}\label{sec:discuss}

In Wang23, we highlight the existence of two kinematic populations across
wide ranges of mass, star formation, and environment, that are distinct in quenching history.
Wang23 shows that the stellar metallicity of slow-rotating galaxies is star formation independent which implies that gas outflows play a pivotal role in quenching their star formation \citep[see also ][]{2015Natur.521..192P}.
This work, enabled by the full data release of MaNGA and its extensive H\textsc{i} follow-up, directly shows that the galaxy population of slower rotation are generally more deficient in cold gas than the fast-rotating counterparts and lends empirical support to a outflow-driven fast quenching history.
The fast-rotating population, on the other hand, are more consistent with a quiescent and starvation-driven quenching history given their significant metal enrichment as star formation goes down.
Their remaining abundant H\textsc{i} reservoir, currently non-star-forming due to, e.g., large angular momentum \citep{2020MNRAS.491L..51P,2022MNRAS.509.2707L}, may still fuel future star formatio when funneled
in, and thus makes the overall quenching a lasting process.

The cold gas deficiency of the slower population is most prominent for massive galaxies with $\mathcal{M}_{\star}>10^{10.8}\mathcal{M}_{\odot}$, i.e. roughly above the typical $M^*$ of Press-Schechter stellar mass function, when the internal "mass quenching" mechanism becomes efficient \citep{2010ApJ...721..193P}.
Galaxies in this mass regime are very likely to host excessively massive central black holes which have experienced rapid growth since the overall potential became deep enough to support intense gas inflow against supernova feedback \citep{2014MNRAS.438.1870D, 2015MNRAS.450.2327Z, 2018MNRAS.481.3118M}.
The tremendous amount of energy output due to black hole growth can substantially heat the intergalactic and circumgalactic medium, and has been widely thought to be responsible for the decline of star formation efficiency among massive galaxies.
This black hole feedback should have been even stronger, cumulatively, in galaxies of slower rotation given their more massive spheroidal structure and the empirical tight relation between such structure and central black hole mass \citep{2013ARA&A..51..511K}, which is also suggested by the state-of-the-art simulations \citep{2022MNRAS.512.5978R, 2022MNRAS.514.5840P}.

Besides more massive central black holes, the feedback processes in the slow-rotating population can impact cold gas at a higher efficiency.
The disturbances (e.g., mergers) that are responsible for the angular momentum loss of slow-rotating galaxies, tend to stir the dense and planar cold gas into a looser and more volume filling form.
At the same time, disturbances also channel cold gas into the center and simultaneously trigger active galactic nuclei \citep{2020A&A...637A..94G} and star formation, from which the feedback energy, particularly the part\footnote{The higher coupling efficiency also holds for stellar feedback, though not being biconical, in slow-rotating galaxies of volume filling gas.
The nonnegligible amount of energy emitted by evolved stars can thus contribute to the even larger gas deficit found in old passive galaxies.} emitted biconically out of black holes, couples better with the volume filling gas than in the case when gas is dense and planar.
Extreme cases in the local Universe are those ultraluminous infrared galaxies produced by mergers.
Strong outflows of cold gas are often associated with these systems and are the main driver of gas depletion \citep[e.g.,][]{2017ApJ...836...11G}.
We do find evidence from the single-dish H\textsc{i} spectra of star-forming slow-rotating galaxies, consistently at all available masses.
Their emission lines are significantly more concentrated at low velocities compared to star-forming fast-rotating galaxies, which is expected as a result of having a massive component of lagging thick gas \citep{2006MNRAS.366..449F} and is reminiscent of the observed connection between mergers and concentrated H\textsc{i} line profiles \citep{2022ApJ...929...15Z}.
Such a picture of elevating feedback efficiency is illustrated in \autoref{fig:draw}.

The absence of the cold gas deficiency among intermediate mass ($\mathcal{M}_{\star}\sim10^{10.5}\mathcal{M}_{\odot}$) slow-rotating galaxies then appears to be plausible along the reasoning.
These galaxies just grow to such critical mass \citep{2019arXiv190408431D} that the gravitational potentials are deep enough to resist supernovae feedback while central black holes are yet to enter the fast growth phase, a possible reason also for the peak star formation efficiency occurring at this mass \citep{2018ARA&A..56..435W}.

We enclose the discussion by returning to the noticeable and unambiguous gap in gas mass between the two kinematic populations at high stellar mass.
Before this work, there were many attempts for catching in the act the negative feedback to gas from the central black holes.
By comparing the cold gas content of galaxies hosting or not an active galactic nucleus, people found no evidence for feedback expelling a large amount of gas \citep[e.g.,][]{2018ApJ...854..158S}.
However, the active phase of central black holes can commonly be as short as one Myr, and the effects are either overall too weak to impact the entire gas reservoir, or the timescale is too short to see any significant change instantly.
By contrast, the integrated effects of black hole feedback can be observed more readily.
Even though slow-rotating galaxies do not host clearly more black hole activities at the moment (as seen in our data using optical line diagnostics), they have more massive spheroidal structure and must have borne stronger feedback from the central black holes across the evolution history as compared to the disc-dominated fast-rotating galaxies.
The cold gas deficiency of massive slower population thus serves as strong evidence, in an accumulative sense, for the negative role of central black holes in the star formation history.

\section{Conclusion}\label{sec:con}

This work compares the cold gas content of galaxies of contrasting intrinsic morphology, in order to find key empirical evidence for the outflow driven quenching history of spheroid-dominated galaxies suggested by our previous work Wang23.
Using the largest sample to date, we robustly measure the spin parameter $\lambda_{R_{\rm e}}$, i.e. the normalized specific angular momentum of stars within one half-light radius, based on IFS data of full MaNGA for about 10000 galaxies, and carefully control the measurement quality to unambiguously classify galaxies by their intrinsic shapes.
Taking non-detections into proper account, we stack the emission line of atomic hydrogen H\textsc{i} to measure the mean H\textsc{i} gas fractions of galaxies at given stellar mass, star formation level, and stellar kinematics, using L-band spectra of H\textsc{i}-MaNGA survey for two thirds of the MaNGA sample.

We find the following results, unchanged when studying central galaxies of comparable dark matter halo mass:
The galaxy population of slower stellar rotation, or equivalently of morphology dominated by spheroidal structure, are generally more H\textsc{i} gas-poor than the disc-dominated population of faster rotation, at the same stellar mass and star formation level.
This cold gas deficit of slow-rotating galaxies is largely suppressed at intermediate mass regime $10^{10.2}<\mathcal{M}_{\star}/\mathcal{M}_{\odot}<10^{10.8}$ and is most significant for massive galaxies of mass $\sim10^{11}\mathcal{M}_{\odot}$ below the star formation main sequence.
From the shape of stacked H\textsc{i} line spectra, star-forming slow-rotating galaxies have more cold gas moving at low velocity along the line of sight compared to that of star-forming fast-rotating galaxies, consistently at all stellar masses.
Such striking contrast of narrow single-peaked versus broad double-peaked spectrum shape, suggests that the cold gas in galaxy population of slower rotation is disturbed and unsettled onto the disc.

Our results provide evidence that gas outflows play an important role in the evolution of slow-rotating galaxies, particularly at high mass, and imply the negative feedback from central black holes given that slow-rotating galaxies on average have more massive central black holes.
One key ingredient in the outcoming picture, hinted at by H\textsc{i} velocity distributions, is that feedback impacts disturbed gas more efficiently.
As mergers destroy both stellar and gas discs, and trigger violent star formation with the central black hole fed by inflowing gas, the active galactic nucleus can heat the disturbed volume filling gas biconically and drives outflows at a mass loading factor that would otherwise be much lower when gas is in a thin plane.
Future works based on high-resolution simulations and radio interferometric data can lead to a better understanding of how the black hole feedback interacts with cold gas of different dynamical states.

\section*{Acknowledgements}

We thank our anonymous referee for the helpful comments.
YP and BW acknowledge the National Key R\&D Program of China Grant 2022YFF0503401, National Science Foundation of China (NSFC) Grant No. 12125301, 12192220, 12192222, and the science research grants from the China Manned Space Project with No. CMS-CSST-2021-A07.
BW wants to thank Shengdong Lu for useful comments and Inkscape's contributors for creating Inkscape as a free, open source, and powerful vector graphics editor.

Funding for the Sloan Digital Sky Survey IV has been provided by the Alfred P. Sloan Foundation, the U.S. Department of Energy Office of Science, and the Participating Institutions. SDSS acknowledges support and resources from the Center for High- Performance Computing at the University of Utah. The SDSS website is \url{www.sdss.org}.

SDSS-IV is managed by the Astrophysical Research Consortium for the
Participating Institutions of the SDSS Collaboration including the
Brazilian Participation Group, the Carnegie Institution for Science,
Carnegie Mellon University, the Chilean Participation Group, the French Participation Group, Harvard-Smithsonian Center for Astrophysics,
Instituto de Astrof\'isica de Canarias, The Johns Hopkins University, Kavli Institute for the Physics and Mathematics of the Universe (IPMU) /
University of Tokyo, the Korean Participation Group, Lawrence Berkeley National Laboratory,
Leibniz Institut f\"ur Astrophysik Potsdam (AIP),
Max-Planck-Institut f\"ur Astronomie (MPIA Heidelberg),
Max-Planck-Institut f\"ur Astrophysik (MPA Garching),
Max-Planck-Institut f\"ur Extraterrestrische Physik (MPE),
National Astronomical Observatories of China, New Mexico State University,
New York University, University of Notre Dame,
Observat\'ario Nacional / MCTI, The Ohio State University,
Pennsylvania State University, Shanghai Astronomical Observatory,
United Kingdom Participation Group,
Universidad Nacional Aut\'onoma de M\'exico, University of Arizona,
University of Colorado Boulder, University of Oxford, University of Portsmouth,
University of Utah, University of Virginia, University of Washington, University of Wisconsin,
Vanderbilt University, and Yale University.

\section*{Data Availability}

Stellar mass and SFR of MaNGA galaxies are taken from GSWLC-X2 catalogue \citep{2016ApJS..227....2S,2018ApJ...859...11S} available on https://salims.pages.iu.edu/gswlc/.
The H\textsc{i} spectra and measurements can be found at https://data.sdss.org/sas/dr17/env/MANGA\_HI/.

Supplementary data are available at ApJL online:

\autoref{tab1}. Table of photometric and kinematic properties for the 9793 MaNGA galaxies in our main sample.


\bibliographystyle{aasjournal}



\end{document}